\pdfoutput=1
\documentclass[journal]{IEEEtran}

\usepackage{amsmath,amssymb,amsfonts}
\usepackage{booktabs}
\usepackage{array}
\usepackage{graphicx}
\usepackage[hidelinks]{hyperref}
\usepackage{xcolor}
\usepackage{enumitem}
\usepackage{textcomp}
\usepackage{url}
\usepackage{multirow}
\begin{document}

\title{Context-Binding Gaps in Stateful Zero-Knowledge Proximity Proofs: Taxonomy, Separation, and Mitigation}

\author{Yoshiyuki Ootani%
\thanks{Yoshiyuki Ootani is an independent researcher based in Japan (e-mail: info@ootanl.com).}}

\maketitle

\begin{abstract}
A zero-knowledge proximity proof certifies geometric nearness but carries no commitment to an application context.
In stateful geo-content systems --- where drops share coordinates, policies evolve, and content has persistent identity --- this gap can allow proof transfer between application objects unless additional operational invariants are maintained.
We present a systems-security analysis of this deployment problem: a taxonomy of context-binding vulnerabilities, a formal off-circuit verification model with a replay game that asks whether a \emph{recorded proof transcript} can be re-bound to a different application context (fresh in-radius proving is provably outside the reach of any statement-level mechanism and is explicitly delegated to an orthogonal presence layer), an assumption comparison across five binding strategy classes, and a concrete instantiation, Zairn-ZKP, that embeds drop identity, policy version, and session context as public circuit inputs.
Compared with a strong stored-digest server-checking alternative, in-proof binding removes the nonce-to-drop mapping and nonce-uniqueness invariants from the operational assumption set --- the association between issued digest values and application objects remains an issuance responsibility for every strategy --- and adds no measurable proving cost relative to the sound geo-only baseline ($-0.12$\,ms median in our setup).
It also removes a correctness pitfall we identify empirically: a plausible off-circuit implementation that omits one server-side check remains vulnerable to cross-drop transfer.
A sensitivity analysis across six network conditions, POI density measurements at seven venues in four countries, and an epoch-window deployment simulation (110--506 vulnerable transfer pairs per 60\,s epoch at the measured urban venues) indicate that, in dense urban deployments, same-epoch transfer is a realistic concern unless per-request nonces are maintained.
When the stored-digest path is hardened it blocks the same transfer attacks under per-request nonces --- but, as we make explicit, its resistance comes from an in-statement challenge digest, making it a hybrid rather than a purely off-circuit design, while purely off-circuit strategies cannot resist an adversary who can request fresh challenges; a controlled comparison holding nonce policy constant shows that in-proof context binding is the only strategy that blocks same-epoch transfer under shared nonces, reduces the assumption surface, and avoids per-drop server state in our design, while nonce policy drives most of the latency difference.
Evaluation across five platforms, seven binding strategies, an implementation comparison (110 vs.\ 20~LOC; six vs.\ one failure modes), and an end-to-end cross-drop transfer attack supports this analysis. The contribution is therefore not a new cryptographic primitive, but a deployable methodology for reducing assumption surfaces in stateful ZK-backed verification workflows; all artifacts are publicly available.
\end{abstract}


\section{Introduction}
\IEEEPARstart{A}{} generic zero-knowledge proximity proof certifies that the prover knows coordinates within a specified radius of a target --- but the proof itself carries no cryptographic commitment to any application context: it is a \emph{free-floating geometric fact}.
Recent work has advanced geometric precision~\cite{ernstberger2024zklp} and integrated session nonces for replay resistance~\cite{lauinger2024portal}.
However, in \emph{stateful geo-content systems} --- where multiple drops may share identical coordinates, policies evolve across epochs, and content has persistent identity --- the gap between geometric proof and application semantics creates vulnerability classes that have not been systematically studied.

Server-side enforcement can close many of these gaps: a hardened stored-digest check (strategy~2c) blocks the same transfer attacks when combined with per-request nonces --- though its resistance, we show, comes precisely from a challenge digest bound \emph{inside} the proof statement, making it a hybrid rather than a purely off-circuit design.
The question is not whether server-side enforcement \emph{can} be made secure, but \emph{which assumptions it requires}, where its binding authority actually lives, and what happens when one assumption fails (\S\ref{sec:off-circuit-model}).

The contribution is not a new cryptographic primitive --- binding values to ZKP public inputs is a known technique~\cite{lauinger2024portal} --- but a \emph{security-engineering analysis} of \emph{which deployment assumptions disappear} when context is cryptographically enforced rather than operationally checked.
Concretely:
\begin{itemize}[leftmargin=*]
  \item[\textbf{C1.}] A \emph{deployment-security methodology} for stateful ZK-backed workflows: a vulnerability taxonomy separating two context-binding gaps (V1, V3) from a circuit-soundness pitfall (V2), plus a formal \emph{off-circuit verification model} and assumption-based comparison that clarifies which checks remain operational and which can be cryptographically enforced (Table~\ref{tab:trust-surface}).
  \item[\textbf{C2.}] \emph{Controlled factor separation} (Table~\ref{tab:same-policy}): holding nonce policy constant across four strategies isolates two independent effects --- binding location drives assumption surface and correctness fragility, while nonce policy drives latency and state --- supported by seven strategies $\times$ seven scenarios, epoch-window deployment simulation (Table~\ref{tab:epoch-vuln}), implementation comparison, end-to-end attack, and five platforms (\S\ref{sec:eval}).
  \item[\textbf{C3.}] \emph{Zairn-ZKP}~\cite{zairn,zairn_zenodo}: a concrete context-bound instantiation (binding level~(iii)) used to realize and evaluate the methodology within a three-layer defense architecture (\S\ref{sec:design}).
\end{itemize}
GPS truthfulness (V4) is an orthogonal trust boundary discussed in \S\ref{sec:v4-discussion}.

\section{Background and Related Work}
\label{sec:background}

\subsection{Location privacy and proof-of-location}
Location privacy has long been recognized as central to pervasive computing~\cite{beresford2003location,krumm2009survey,chatzikokolakis2015geo}.
Proof-of-location (PoL) systems address location verification: Saroiu and Wolman~\cite{saroiu2009enabling} formalized location proofs; APPLAUS~\cite{zhu2011applaus} and VeriPlace~\cite{luo2010veriplace} introduced privacy-preserving variants; STAMP~\cite{wang2016stamp} addressed mobile contexts.
More recently, Akand et al.\ presented composable anonymous PoL~\cite{akand2023composable} and geo-tampering-resistant designs~\cite{akand2023geotamper}.
These systems focus on issuance protocols, anonymity, and non-transferability rather than on application-bound unlock semantics within a deployable software stack.

\subsection{Zero-knowledge proof systems}
Zero-knowledge proofs build on foundational work by Goldwasser et al.~\cite{goldwasser1989knowledge} and practical SNARK constructions including Pinocchio~\cite{parno2013pinocchio}, Groth16~\cite{groth16}, and PLONK~\cite{gabizon2019plonk}.
The Circom circuit description language~\cite{bellesmunoz2023circom,circom} and the snarkjs runtime~\cite{snarkjs} have made Groth16-based applications practical for web and mobile deployment.
Recent work on ZK circuit security~\cite{wen2023practical,liu2023coda,stephens2025consistency,pailoor2023underconstrained} has shown that underconstrained circuits are a recurring source of deployed vulnerabilities.

For location privacy, ZKLP~\cite{ernstberger2024zklp} demonstrated accurate floating-point SNARKs for proximity checks but does not address application-layer binding.
Portal~\cite{lauinger2024portal} proposed time-bound, replay-resistant ZKP verification for single sign-on; token binding and channel binding address analogous out-of-context replay in web authentication~\cite{popov2018tokenbinding,bhargavan2015channelbindings}.
Blockchain-based ZK-PoL has also been explored~\cite{bensamuel2020zkpol}.

\subsection{GPS spoofing and location privacy attacks}
GPS spoofing is practically feasible~\cite{humphreys2008spoofing,tippenhauer2011spoofing}; countermeasures span RAIM~\cite{brown1992raim}, sensor fusion~\cite{spens2022android,dasgupta2022sensor}, cellular cross-referencing~\cite{oligeri2019cellular}, and hardware TEEs~\cite{liu2012trustedsensors}.
Inference attacks can deanonymize obfuscated location data~\cite{krumm2007inference,wernke2014classification}, motivating zero-knowledge approaches (\S\ref{sec:baseline-eval}).

\subsection{Positioning this work}
Three research threads can be distinguished:
\emph{(i)}~PoL/LPA systems address issuance, anonymity, and geo-tampering resistance~\cite{saroiu2009enabling,luo2010veriplace,akand2023composable,akand2023geotamper}.
\emph{(ii)}~Geospatial ZK systems address geometric accuracy and efficient private proximity testing~\cite{ernstberger2024zklp}.
\emph{(iii)}~This paper addresses \emph{context-binding vulnerabilities}: identifying, classifying, and mitigating the gaps that appear when a generic proximity proof is deployed in a stateful geo-content system.
Binding values to ZKP public inputs is a known technique: Portal~\cite{lauinger2024portal} binds session nonces for replay resistance.
Our contribution is the identification and systematic study of \emph{geo-content-specific} failure modes that nonce binding alone does not address.

\textbf{Binding levels.}
We distinguish three levels of binding between a ZK proof and its deployment context:
\emph{(i)}~\emph{off-circuit nonce check} --- a session nonce is verified alongside but outside the proof (no public-input commitment);
\emph{(ii)}~\emph{in-proof session nonce} --- the nonce is a public circuit input, preventing temporal replay (Portal~\cite{lauinger2024portal});
\emph{(iii)}~\emph{in-proof application context} --- drop identity, policy version, and session nonce are all public circuit inputs (Zairn-ZKP).
Level~(ii) prevents cross-session proof reuse; level~(iii) additionally prevents cross-drop transfer within the same freshness epoch and cryptographically commits to application-semantic properties that level~(ii) does not capture (\S\ref{sec:off-circuit-model}, \S\ref{sec:off-circuit-eval}).
Table~\ref{tab:comparison} summarizes how this work relates to the closest systems.

\begin{table*}[t]
  \centering
  \renewcommand{\arraystretch}{1.15}
  \caption{Comparison with related systems. ``Context bind.'' indicates whether the proof cryptographically commits to application context.}
  \label{tab:comparison}
  \footnotesize
  \begin{tabular}{lcccccc}
    \toprule
    System & Proof system & Precision & Context bind. & Replay resist. & GPS integrity & Open source \\
    \midrule
    APPLAUS~\cite{zhu2011applaus} & None (PoL) & GPS-level & N/A & Partial & Not addressed & No \\
    STAMP~\cite{wang2016stamp} & None (PoL) & GPS-level & N/A & Yes & Not addressed & No \\
    Akand et al.~\cite{akand2023composable} & Composable PoL & GPS-level & N/A & Yes & Not addressed & No \\
    ZKLP~\cite{ernstberger2024zklp} & Custom SNARK & Float & No & No & Not addressed & Partial \\
    Portal~\cite{lauinger2024portal} & Groth16 & Fixed-pt. & Session nonce$^{*}$ & Yes (time) & Not addressed & No \\
    \textbf{Zairn-ZKP (this work)} & Groth16 & Fixed-pt. & \textbf{App.\ context}$^{\dagger}$ & \textbf{Yes} & Partial / heuristic & \textbf{Yes} \\
    \bottomrule
    \multicolumn{7}{l}{\footnotesize $^{*}$Binding level~(ii): session nonce as public input prevents temporal replay but not cross-context transfer.}\\
    \multicolumn{7}{l}{\footnotesize $^{\dagger}$Binding level~(iii): drop identity, policy version, epoch, and session nonce as public inputs.}
  \end{tabular}
\end{table*}

\section{Threat Model and Security Goals}
\label{sec:threat}

\subsection{System roles}
Zairn involves four roles: a drop creator, a drop claimant, the application backend, and optional persistence services (IPFS, EVM registry).
The backend manages authentication, database access, policy enforcement, and claim recording.
The claimant proves proximity to unlock encrypted content.

\subsection{Adversary model}
We consider adversaries that can:
\begin{itemize}[leftmargin=*]
  \item observe or modify network traffic to the extent allowed by the deployment channel;
  \item submit malformed proof payloads, replay captured proofs, or attempt cross-drop reuse;
  \item spoof GPS coordinates via software (mock location APIs) or hardware (SDR-based GPS simulators);
  \item use VPN or proxy services to mask network-derived geolocation;
  \item control the client application (e.g., modified app binary that bypasses off-circuit checks).
\end{itemize}

We do not claim protection against fully compromised client hardware with root-level sensor access, coercive physical attacks, or global traffic analysis for unlinkability.

\subsection{GPS integrity assumptions}
\label{sec:gps-integrity}
A critical trust boundary exists at the location input: the ZKP path proves proximity of \emph{stated} coordinates, but the system cannot cryptographically verify that these coordinates reflect the prover's actual physical position.
This gap (V4) is inherent to all client-generated location proofs and is discussed in \S\ref{sec:discussion}.

\subsection{Security goals}
\begin{itemize}[leftmargin=*]
  \item \textbf{Content confidentiality:} encrypted payloads remain unreadable without the correct decryption context.
  \item \textbf{Proximity soundness:} an accepted proof corresponds to private coordinates satisfying the proximity relation.
  \item \textbf{Context binding:} a proof for one drop or session is not reusable for another.
  \item \textbf{Replay resistance:} stale or captured proofs are rejected once their bound context expires.
  \item \textbf{Coordinate privacy:} the verifier learns statement validity without learning exact coordinates.
\end{itemize}
As an artifact goal (not a security property), we also target \emph{reproducibility}: the full stack is inspectable, buildable, and benchmarkable from public artifacts (\S\ref{sec:repro}).

\subsection{Defense-in-depth rationale}
\begin{figure}[t]
  \centering
  \includegraphics[width=\linewidth]{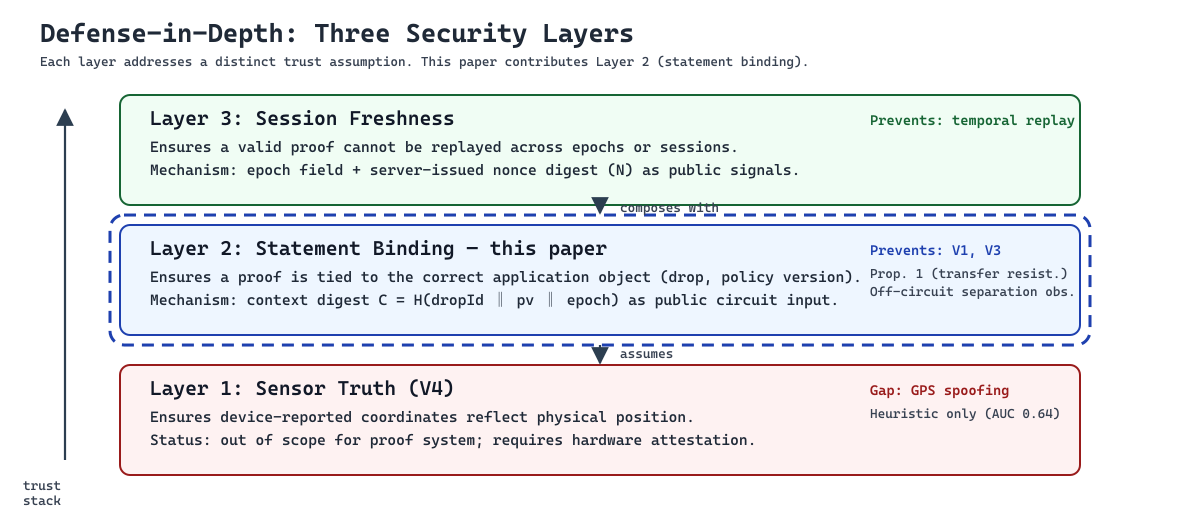}
  \caption{Three-layer defense-in-depth. This paper contributes Layer~2 (statement binding). Layer~1 (sensor truth) is out of scope; Layer~3 (session freshness) composes with Layer~2 via the epoch and nonce fields.}
  \label{fig:three-layers}
\end{figure}

The system should be read as three layers with distinct responsibilities.
\emph{Sensor truth} concerns whether device-reported coordinates reflect physical position; this is the V4 gap and is not solved by the proof system.
\emph{Statement binding} concerns whether a proof is tied to the correct application object (drop, policy version); this is the V1/V3 gap addressed by context-bound public signals.
\emph{Session freshness} concerns whether an otherwise valid proof can be replayed across epochs or sessions; this is addressed by the epoch and nonce fields.
A zero-knowledge proximity proof covers private geometry, not sensor truth.
The contribution of this paper is therefore the middle layer --- statement binding in a stateful deployment --- together with an explicit account of how it composes with freshness and with orthogonal sensor-attestation mechanisms.
Figure~\ref{fig:three-layers} illustrates these layers.

\subsection{Off-circuit verification model}
\label{sec:off-circuit-model}
To make claims about off-circuit strategies precise, we define the class of \emph{off-circuit verification} as follows.
An off-circuit verifier receives a proof $\pi$, its public signals $\mathit{pub}$, and an auxiliary claim $\mathit{aux}$ (which may include signed tokens, stored digests, or any authenticated metadata).
The verifier:
(a)~checks $\mathsf{Groth16.Verify}(vk, \mathit{pub}, \pi) = 1$;
(b)~applies an arbitrary (possibly cryptographic) predicate $P(\mathit{pub}, \mathit{aux})$;
but \emph{does not} require that $\mathit{aux}$ is committed inside $\pi$.
The key structural constraint is: since $\pi$ was generated by a generic proximity circuit (containing only geometric inputs), no information about application context is encoded in $\mathit{pub}$.
Predicate $P$ can therefore validate $\mathit{aux}$ against external state (database, signed tokens) but cannot compare $\mathit{aux}$ against the proof's internal commitment, because no such commitment exists.
This model encompasses client-resident checks, server-recomputed digests, stored-digest lookups, and authenticated-challenge protocols.

\textbf{Observation (Off-circuit assumption gap).}
Let $\mathcal{C}$ be any ZK circuit whose public signals encode only the attested property $\mathit{prop}$ (e.g., geometric parameters), not the identity of the application object being attested.
For any off-circuit predicate $P(\mathit{pub}, \mathit{aux})$, if the adversary can legitimately obtain $\mathit{aux}_2$ valid for object~$o_2$, then a proof generated for $o_1$ (with $\mathit{prop}_1 = \mathit{prop}_2$) will also be accepted by $P$ under~$o_2$, because $\mathcal{C}$ encodes no object identity in $\mathit{pub}$.
This observation is general: it applies to any ZK circuit that attests a property of a persistent application object --- anonymous credentials, access-control tokens~\cite{popov2018tokenbinding,bhargavan2015channelbindings}, multi-tenant ZKP services --- not only proximity proofs.

The ``if'' condition --- whether the adversary can obtain $\mathit{aux}_2$ --- is the crux.
In geo-content systems, legitimate users routinely request challenges for multiple drops at the same location, so $\mathit{aux}_2$ acquisition is straightforward (the adversary simply requests drop~B's challenge); our comparison games therefore grant the adversary a \emph{challenge oracle} that issues fresh, correctly formed challenges for any context (Definition~2).
Against such an adversary, a \emph{purely} off-circuit predicate cannot prevent transfer at all: the Observation applies verbatim, because no acceptance condition ties the recorded transcript to any particular $\mathit{aux}$ --- a freshly issued token or recomputed digest for drop~B pairs cleanly with a replayed proof from drop~A.
When a deployed ``off-circuit'' design does resist transfer, it is because a challenge value has been moved \emph{into the statement}: the stored-digest strategy we implement (2c-hardened, \S\ref{sec:impl-complexity}) has the client place the issued challenge digest in a public signal and the server compare that signal against stored state, a \emph{hybrid} in which session binding is in-statement while drop-identity binding remains an operational invariant (nonce-to-drop mapping, per-request nonce uniqueness).
The observation thus identifies both the structural limit of purely off-circuit enforcement and the \emph{assumption set} on which hybrid correctness depends (Scenarios~F--G in \S\ref{sec:off-circuit-eval}; implementation evidence in \S\ref{sec:impl-complexity}).

In-proof context binding (\S\ref{sec:design}) takes a different approach: embedding application context in $\mathit{pub}$ makes the proof itself context-dependent ($\mathit{pub}_1 \neq \mathit{pub}_2$), moving drop-identity and policy-version binding from post-issuance operational invariants to cryptographic enforcement (the transcript is pinned to the issued digest values; \S\ref{sec:off-circuit-eval} makes the accounting precise).
Table~\ref{tab:trust-surface} in \S\ref{sec:off-circuit-eval} compares the full assumption set across five strategy classes; Comparison Result~1 summarizes the design-space consequences of different operational and cryptographic assumptions.

\section{Vulnerability Analysis}
\label{sec:vuln}

This section presents a systematic classification of vulnerabilities that arise when a generic zero-knowledge proximity proof is deployed in a stateful geo-content unlock system.
The \emph{primary axis} of the taxonomy is V1 and V3: \emph{context-binding} vulnerabilities where the gap between proof content and application semantics enables proof transfer.
These form the core problem that this paper addresses; the assumption-aware strategy comparison (Comparison Result~1) and sensitivity analysis (Table~\ref{tab:sensitivity}) are structured around them.
V2 is an independent \emph{circuit-soundness} pitfall included because the same underconstrained-comparison pattern recurs across deployed Circom projects~\cite{wen2023practical,liu2023coda,pailoor2023underconstrained}; it supports deployable soundness, but is orthogonal to the main binding question.
V4 is a \emph{sensor-trust} boundary outside the proof system's scope.
Each vulnerability class is illustrated with a concrete attack scenario and its consequences.

\subsection{V1: Unbound statement (cross-drop replay)}
A generic proximity circuit proves ``the prover's coordinates are within radius $r$ of target $(lat_t, lon_t)$.''
If two drops share the same target and radius, a proof generated for drop~A is valid for drop~B.
Even if targets differ, an adversary who generates a proof while physically present at one location can replay it later against any drop at that location.
Without context binding, the proof attests only to geometric proximity, not to a specific unlock session.

\textbf{Boundary with V4.}
V1 concerns \emph{proof-transcript} reuse: the same proof $(\pi, \mathit{pub})$ is submitted to a different context.
A distinct threat --- an adversary who feeds valid in-radius coordinates to the prover to generate a \emph{fresh} proof (no prior visit is even required: the target coordinates are public) --- is a V4 (GPS truthfulness) issue: the system receives genuine-looking coordinates that do not reflect physical presence.
Context binding (Proposition~1) prevents the former; only location attestation can address the latter.

\textbf{Consequence:} cross-drop content theft and temporal replay.

\subsection{V3: Application-layer binding fragility}
An intermediate design binds context off-circuit: client code checks that the proof's public signals match the expected drop parameters.
An adversary who controls the client (modified binary, DevTools, proxy rewriting) bypasses this check entirely.
A server-side off-circuit check is not subject to this bypass, but still relies on the backend correctly implementing and synchronizing validation logic; in-proof binding eliminates this dependency by making context part of the mathematical statement.

\textbf{Consequence:} client-resident off-circuit binding provides no security against a client-side adversary; server-side binding is more robust but introduces an operational trust dependency.
We evaluate seven binding strategies in \S\ref{sec:off-circuit-eval}.

\subsection{Supporting findings: V2 and V4}
The following findings are orthogonal to V1/V3 but affect practical security of proximity-proof deployments.

\textbf{V2: Unconstrained comparison (supporting circuit-soundness pitfall).}
In BN128, $r^2 - d^2$ wraps to a large field element when $d > r$; without bit-decomposition, R1CS constraints remain satisfiable for out-of-range witnesses~\cite{wen2023practical,liu2023coda}.
In the prototype, the unconstrained hint \texttt{out <-- (r2 >= dist2) ? 1 : 0} allows the prover to assign \texttt{out = 1} at any distance.
\emph{Consequence:} proof generation succeeds beyond the configured radius.

\textbf{V4: Assumed GPS truthfulness (sensor-trust boundary).}
GPS spoofing is feasible with commodity hardware~\cite{humphreys2008spoofing}; a spoofer generates a \emph{cryptographically valid} proof at a fabricated location.
This is outside the proof system's scope: ZK guarantees coordinate-relation satisfaction, not physical accuracy~\cite{psiaki2016gnss}.
Context binding remains necessary regardless of V4 resolution: even with perfect GPS, a proof for drop~A is replayable to drop~B unless it binds drop identity.
\emph{Consequence:} GPS spoofing bypasses ZKP verification entirely.

\subsection{Summary}
Table~\ref{tab:vuln-summary} summarizes the four vulnerability classes and their mitigations.

\begin{table*}[t]
  \centering
  \renewcommand{\arraystretch}{1.15}
  \caption{Vulnerability taxonomy for zero-knowledge proximity proofs in stateful applications.}
  \label{tab:vuln-summary}
  \begin{tabular}{cllllc}
    \toprule
    ID & Category & Vulnerability & Attack vector & Mitigation & Section \\
    \midrule
    \multicolumn{6}{l}{\emph{Primary: context-binding gaps (this paper's core contribution)}} \\
    V1 & Context binding & Unbound statement & Cross-drop replay & In-proof context binding & \S\ref{sec:design} \\
    V3 & Context binding & App-layer binding fragility & Client-resident bypass & In-proof binding & \S\ref{sec:design} \\
    \midrule
    \multicolumn{6}{l}{\emph{Supporting: orthogonal to context binding}} \\
    V2 & Circuit soundness & Unconstrained comparison & Boundary inflation & Bit-decomposed \texttt{LessEqThan} & \S\ref{sec:design} \\
    V4 & Sensor trust & Assumed GPS truthfulness & GPS spoofing & Heuristic scorer$^{*}$ & \S\ref{sec:v4-discussion} \\
    \bottomrule
    \multicolumn{6}{l}{\footnotesize $^{*}$Partial mitigation only; hardware attestation needed for strong guarantees (\S\ref{sec:v4-discussion}).}
  \end{tabular}
\end{table*}

\section{Zairn-ZKP Design}
\label{sec:design}

\subsection{Design principles}
Zairn-ZKP is guided by three principles:
\emph{sound statement design} (constraints are enforced within the circuit, not delegated to application code),
\emph{context binding} (the accepted statement is specific to the intended drop and session),
and \emph{operational compatibility} (the proof flow fits a realistic application stack).

\subsection{Statement shape}
The application-level goal is that the prover demonstrates:
\begin{equation}
\begin{aligned}
\text{``I know } (lat_u, lon_u) \text{ within radius } r \text{ of } (lat_t, lon_t), \\
\text{for drop } \mathit{dropId} \text{ under policy } \mathit{pv} \text{ at epoch } \mathit{epoch} \text{.''}
\end{aligned}
\end{equation}

\textbf{Implemented circuit relation.}
The circuit does \emph{not} hash preimages internally; instead, it binds to digest \emph{values} computed by the backend at issuance.
The backend computes $C = H(\mathit{dropId} \| \mathit{pv} \| \mathit{epoch})$ and $N = H(\mathit{nonce})$, where $H(x) = \mathsf{SHA\text{-}256}(x) \bmod p$ ($p$ is the BN128 scalar field prime, $p \approx 2^{254}$).
The circuit relation $\mathcal{R}$ accepts $(pub, w)$ iff:
\begin{equation}
\label{eq:circuit-relation}
\begin{aligned}
& d_{\cos\phi}\bigl((lat_u, lon_u), (lat_t, lon_t)\bigr)^2 \le r^2 \\
& \wedge\; pub = (1,\; lat_t,\; lon_t,\; r^2,\; \cos\phi,\; C,\; \mathit{epoch},\; N)
\end{aligned}
\end{equation}
where $w = (lat_u, lon_u)$ is the private witness (claimant coordinates).
The circuit enforces proximity; the \emph{semantic link} between digest values and application context depends on correct backend issuance (Proposition~1, assumption~(ii)).

\textbf{Public signal ordering.}
Groth16 exposes 8 public signals in the order:
\[
\mathit{pub} = (\underbrace{\mathit{out}}_{[0]},\; \underbrace{lat_t, lon_t, r^2, \cos\phi}_{\mathit{geo}\;[1..4]},\; \underbrace{C, \mathit{epoch}, N}_{\mathit{ctx}\;[5..7]})
\]
where $\mathit{out} = 1$ is the proximity result bit (constrained to~1 by the circuit).
We write $\mathit{pub}.\mathit{geo}$ for signals $[1..4]$ and $\mathit{pub}.\mathit{ctx}$ for signals $[5..7]$.

\textbf{Hash-to-field.}
Inputs to $H$ use length-prefixed canonical encoding: each field is preceded by a four-digit decimal byte-length, then concatenated (e.g., \texttt{"0007drop-42000120001\allowbreak 7"}). This provides formally unambiguous domain separation regardless of field content, without requiring separator-character restrictions.
Reducing a 256-bit hash output modulo the ${\approx}$254-bit prime does \emph{not} yield a near-uniform field element: $2^{256} \approx 5.29\,p$, so field elements have five or six 256-bit preimages and the statistical distance of the reduced output from uniform is on the order of several percent.
The analysis never relies on uniformity of $H$'s output --- only on its collision resistance.
The collision resistance of $H$ is inherited from SHA-256 modulo a small additional term: any collision in $H$ implies either a collision in SHA-256 or two distinct 256-bit outputs congruent mod the 254-bit prime~$p$; the latter event has probability $\le q^2/2p$ for $q$ queries ($q \le 2^{64}$), which is negligible but strictly weaker than raw SHA-256 collision resistance.
Proposition~1's assumption~(i) therefore reduces to SHA-256 collision resistance plus this negligible mod-$p$ term.
The circuit operates over BN128 (alt-bn128), the de-facto target for snarkjs/circom tooling, providing $\sim$100-bit security~\cite{groth16}.
This addresses V1 (unbound statement) by making proof validity conditional on both proximity and the currently active application context.

\subsection{Public and private inputs}
Table~\ref{tab:zairn-zkp-inputs} summarizes the statement interface.

\begin{table*}[t]
  \centering
  \renewcommand{\arraystretch}{1.15}
  \caption{Zairn-ZKP statement inputs.}
  \label{tab:zairn-zkp-inputs}
  \begin{tabular}{p{2.6cm}p{1.5cm}p{7.8cm}}
    \toprule
    Input & Visibility & Purpose \\
    \midrule
    $lat_t, lon_t$ & public & Target location of the unlock statement. \\
    \addlinespace[2pt]
    $r^2$ & public & Squared unlock radius used by the verifier and the circuit. \\
    \addlinespace[2pt]
    $\cos(\phi) \times 10^6$ & public & Cosine-latitude scaling factor for longitude correction. \\
    \addlinespace[2pt]
    $C$ & public & Context digest binding the proof to a drop and policy version. \\
    \addlinespace[2pt]
    $epoch$ & public & Freshness window identifier for expiring older statements. \\
    \addlinespace[2pt]
    $N$ & public & Digest of a server-issued nonce for session-level replay resistance. \\
    \addlinespace[2pt]
    $lat_u, lon_u$ & private & Claimant coordinates, hidden from the verifier. \\
    \bottomrule
  \end{tabular}
\end{table*}

\subsection{Context binding and freshness}
Context binding addresses V1 directly.
The context digest $C = H(\mathit{LP}(\mathit{drop\_id}, \mathit{policy\_version}, \mathit{epoch}))$, where $\mathit{LP}$ is length-prefixed canonical encoding, binds the proof to the persistent identity and policy of the drop, while the challenge digest $N = H(\mathit{server\_nonce})$ binds it to a specific unlock session.
Even if two drops share the same coordinates and radius, a proof for one context does not verify for the other.
The epoch appears both inside $C$ and as an independent public signal: its role in $C$ is to bind the proof to a specific policy state, while its role as a separate signal allows the verifier to enforce freshness policy (e.g., reject epoch $< \mathit{current} - \delta$) without recomputing the digest.

\textbf{Adversary classes.}
A \emph{fresh-proving adversary} runs the prover itself.
Because the proving key is public and the proximity relation's witnesses are computable from the public statement alone --- the target coordinates themselves lie within every radius --- \emph{any} party can generate a fresh, cryptographically valid proof for any context; whether the coordinates it feeds the prover reflect physical presence is the sensor-truth question (V4, Layer~1), which no statement-level mechanism can answer.
A \emph{replay adversary} holds a recorded transcript $(\pi_1, \mathit{pub}_1)$ from a prior session and attempts to have \emph{that transcript} accepted for a different context.
Statement binding (Layer~2) is responsible for exactly the second threat, and the guarantee below is scoped accordingly; it says nothing about fresh proving, which it reduces to and delegates to Layer~1 (see the Remark).

\textbf{Definition 1 (Context re-binding game).}
The game $\mathsf{Game}_{\mathcal{A}}^{\mathit{rebind}}(\lambda)$ between a challenger~$\mathcal{C}$ and a PPT adversary~$\mathcal{A}$ proceeds as follows.
\begin{enumerate}[leftmargin=*,itemsep=1pt]
  \item \emph{Setup.} $\mathcal{C}$ runs Groth16 key generation for the Zairn-ZKP relation~$\mathcal{R}$, producing $(pk, vk)$, and gives $(pk, vk)$ to $\mathcal{A}$.
  \item \emph{Two-context issuance.} $\mathcal{C}$ chooses two context tuples $\mathit{ctx}_1, \mathit{ctx}_2$ with $\mathit{ctx}_1 \neq \mathit{ctx}_2$ but $\mathit{geo}_1 = \mathit{geo}_2 = \mathit{geo}$ (identical geometric parameters). For $i \in \{1,2\}$, $\mathcal{C}$ computes canonical digests $C_i = H(\mathit{LP}(\mathit{dropId}_i, \mathit{pv}_i, \mathit{epoch}_i))$, $N_i = H(\mathit{nonce}_i)$, and stores acceptance records $\mathit{rec}_i = (\mathit{geo}, C_i, \mathit{epoch}_i, N_i)$.
  \item \emph{Source transcript.} $\mathcal{C}$ picks a valid proximity witness $w$ and generates an honest proof $\pi_1 \leftarrow \mathsf{Prove}(pk, \mathit{pub}_1, w)$ where $\mathit{pub}_1 = (1, \mathit{geo}, C_1, \mathit{epoch}_1, N_1)$. $\mathcal{C}$ gives $(\pi_1, \mathit{pub}_1)$ and both records to~$\mathcal{A}$.
  \item \emph{Adversary output.} $\mathcal{A}$ outputs auxiliary data $\mathit{aux}^*$ and a claim naming the target record $\mathit{rec}_2$; the submitted transcript is fixed to $(\pi_1, \mathit{pub}_1)$.
  \item \emph{Target acceptance.} $\mathcal{C}$ accepts iff: (a)~$\mathsf{Groth16.Verify}(vk, \mathit{pub}_1, \pi_1) = 1$; (b)~$\mathit{pub}_1.\mathit{geo} = \mathit{geo}$; and (c)~$\mathit{pub}_1.\mathit{ctx} = (C_2, \mathit{epoch}_2, N_2)$.
\end{enumerate}
$\mathcal{A}$ wins if $\mathcal{C}$ accepts. The scheme is \emph{re-binding resistant} if $\Pr[\mathcal{A}\text{ wins}] \le \mathsf{negl}(\lambda)$.

Fixing the submitted transcript is what confines the game to the replay threat; it is a deliberate modeling choice, justified in the Remark below, not a convenience.
The game captures the Scenario~F/G threat: the adversary previously generated a valid proof for drop~1 (the ``source'') while at the location, and later attempts to have \emph{that recorded transcript} accepted for drop~2 (the ``target'') at the same location.

\textbf{Proposition 1 (Replay rejection).} \emph{Under (i)~collision resistance of $H$ and (ii)~correct backend issuance (each $\mathit{rec}_i$ matches the context used at issuance), Zairn-ZKP is re-binding resistant: a transcript accepted for $\mathit{rec}_1$ is rejected for every $\mathit{rec}_2 \neq \mathit{rec}_1$.}

\emph{Proof.}
Acceptance for $\mathit{rec}_2$ requires $\mathit{pub}_1.\mathit{ctx} = (C_2, \mathit{epoch}_2, N_2)$, while by issuance correctness $\mathit{pub}_1.\mathit{ctx} = (C_1, \mathit{epoch}_1, N_1)$.
Since $\mathit{ctx}_1 \neq \mathit{ctx}_2$, the canonical length-prefixed encodings differ (LP is injective), so $C_1 = C_2$ or $N_1 = N_2$ on differing preimages would exhibit a collision in~$H$.
Hence condition~(c) fails except with collision probability, independent of $\mathit{aux}^*$.
\hfill$\square$

\emph{Remark (scope: why fresh proofs are not --- and cannot be --- covered).}
Definition~1 fixes the submitted transcript; it deliberately does not quantify over adversarially generated proofs.
A game that allowed $\mathcal{A}$ to output an arbitrary pair $(\pi^*, \mathit{pub}^*)$ would be unwinnable by \emph{any} scheme for a proximity relation of this shape: $pk$ is public, and a valid witness for the target statement is computable from the public statement itself (choosing $(lat_u, lon_u) = (lat_t, lon_t)$ gives distance zero), so the adversary could always produce a fresh accepting proof for the target context.
Knowledge soundness offers no handle against this --- it guarantees only that an accepting prover knows \emph{some} witness, not a challenger-chosen one.
Nor does transcript malleability add adversarial power worth modeling separately: any accepting pair $(\pi^*, \mathit{pub}_2)$ is by definition a valid proof of the target statement, whose existence is already implied by public proving.
The deployable content of context binding is therefore exactly Proposition~1: \emph{every} transfer attack is reduced to fresh proving for the target statement.
Whether a physically absent party can feed in-radius coordinates to the prover is the Layer-1 presence question (V4), addressed by sensor attestation and orthogonal countermeasures (\S\ref{sec:v4-discussion}), not by any statement-level mechanism.

\emph{Note.}
Without context fields in $\mathit{pub}$, condition~(c) collapses and replay succeeds (\S\ref{sec:off-circuit-model}); this is the attack context binding directly prevents.
If assumption~(ii) fails (e.g., the backend issues $C_1 = C_2$ for distinct contexts), replay succeeds despite in-proof binding; re-binding resistance is a \emph{composed} system property.

This formalizes V1 mitigation; drift resilience follows from digest comparison at issuance rather than recomputation (Table~\ref{tab:drift}).

\subsection{Soundness-oriented circuit redesign}
The Zairn-ZKP circuit addresses V2 through four changes: bounded-domain range checks (\texttt{Num2BitsBounded}), \texttt{LessThan}-gated absolute differences, constrained integer division for cosine scaling, and bit-decomposed final comparison --- aligned with findings on underconstrained ZK circuits~\cite{wen2023practical,liu2023coda,stephens2025consistency,pailoor2023underconstrained}.
Each template is self-contained; the boundary sweep (Table~\ref{tab:boundary-sweep}) empirically cross-checks these guarantees.

\subsection{In-proof vs.\ off-circuit binding}
Embedding context binding in the circuit addresses V3: a client-resident check can be bypassed, whereas in-proof binding makes mismatch detectable by \emph{any} verifier.
Server-side off-circuit checks are more robust and a stored-digest variant reduces operational coupling; the remaining gap is evaluated as Scenario~F in \S\ref{sec:off-circuit-eval}.

\subsection{Integration into the unlock workflow}
The Zairn-ZKP path sits inside the existing GeoDrop unlock workflow.
A claimant requests the public unlock context; the backend returns target parameters, context digest, freshness value, and challenge digest.
The claimant generates a Groth16 proof locally and submits it with public signals.
The verifier checks signal consistency, cryptographic validity, and freshness, then records the claim.

\begin{figure}[t]
  \centering
  \includegraphics[width=\linewidth]{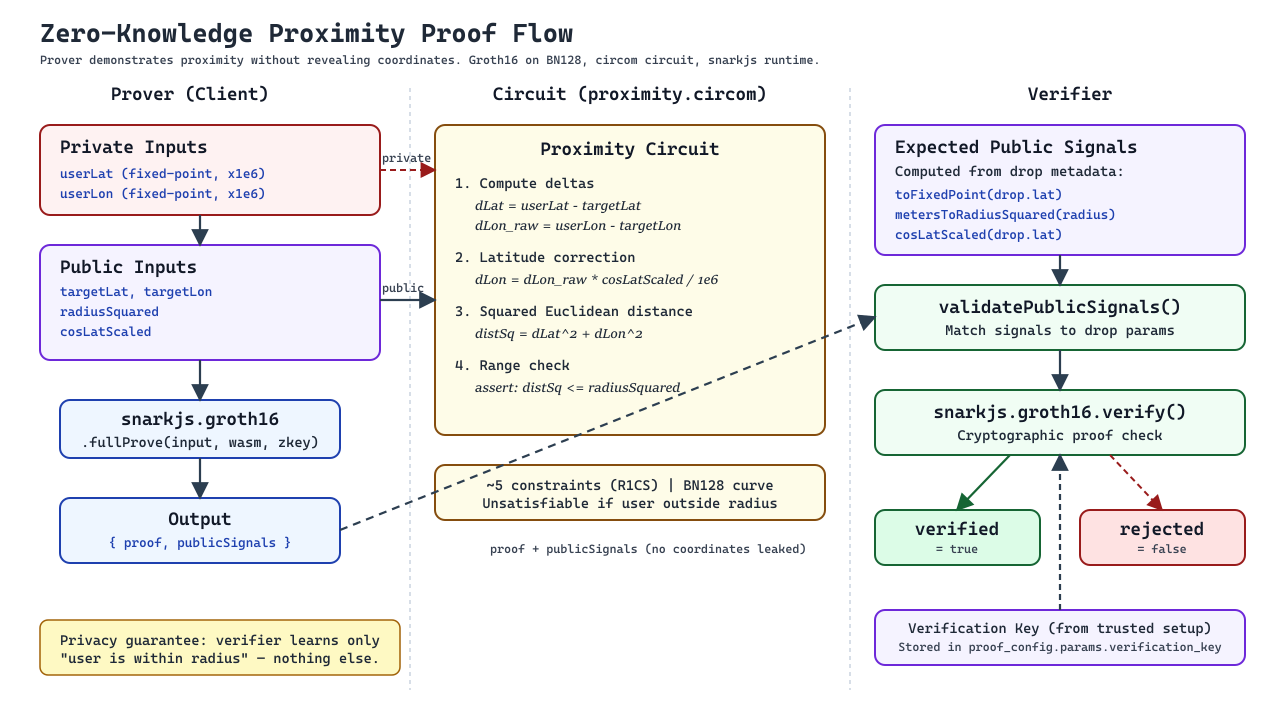}
  \caption{Context-bound Zairn-ZKP unlock flow.}
  \label{fig:zairn-zkp-flow}
\end{figure}

\section{Implementation}
\label{sec:impl}
The Circom circuit (417 non-linear + 57 linear constraints) declares 7 public inputs and 2 private inputs; Groth16 exposes 8 public signals in canonical order $\mathit{pub} = (\mathit{out}, \mathit{geo}[1..4], \mathit{ctx}[5..7])$.
The TypeScript ZKP module exposes \texttt{generateZairnZkpProof} and \texttt{verifyProximityProof}; the verification engine accepts proofs only when both cryptographic validity and public-signal agreement hold.
The implementation resides in the Zairn monorepo~\cite{zairn,zairn_zenodo}.

\section{Evaluation}
\label{sec:eval}

\subsection{Research questions}
\begin{itemize}[leftmargin=*]
  \item \textbf{RQ1 (Integration):} Can the mitigations be integrated into an existing OSS stack without breaking the deployable workflow?
  \item \textbf{RQ2 (Security):} To what extent do the mitigations prevent the identified vulnerability classes?
  \item \textbf{RQ3 (Performance):} What proof-generation and verification overheads are introduced?
  \item \textbf{RQ4 (Binding strategies):} How do no-binding, off-circuit binding, and in-proof binding compare under adversarial scenarios?
  \item \textbf{RQ5 (Baselines):} How does ZKP verification compare with GPS and hash-based alternatives in privacy, cost, and attack resistance?
\end{itemize}

\subsection{Evaluation setup}
Results were collected on five platforms: Machine~A (Windows, x86-64, Node.js v24), Machine~B (Apple M3, macOS, Node.js v22), desktop Chrome WASM, iOS Safari WASM, and Android Chrome WASM.
Desktop benchmarks used $n=50$ runs; mobile $n=10$. All use revision \texttt{1eb05d2} or later.

\subsection{RQ1: Integration feasibility}
The Zairn-ZKP circuit compiles, generates witnesses, produces Groth16 proofs, and verifies them within the existing GeoDrop unlock workflow.
All 98 existing tests pass after integration, confirming non-regression.
The context-bound path coexists with the legacy 5-signal path through backward-compatible validation; security properties are evaluated separately (\S\ref{sec:security-eval}).

\subsection{RQ2: Security improvement}
\label{sec:security-eval}

\textbf{Context-binding ablation.}
Table~\ref{tab:ablation} shows the ablation experiment: a valid proof is resubmitted with one verifier-side context element mutated.
The Groth16 proof remains mathematically valid (it was generated with the original signals), but the signal-validation layer recomputes the expected context digest from the mutated element and rejects the mismatch.
The prototype, lacking context fields, accepts all mutations.

\textbf{Boundary sweep.}
Table~\ref{tab:boundary-sweep} reports a sweep at 50\,m radius (0.5\,m increments, 4 bearings, 2 sites): the prototype accepted all distances 48--52\,m; Zairn-ZKP rejected $>$50\,m consistently.

\textbf{Geometric approximation error.}
Table~\ref{tab:geo-accuracy} reports maximum absolute error across 960 test points (8~latitudes $\times$ 5~radii $\times$ 8~directions $\times$ 3~boundary offsets of $-1$/$0$/$+1$\,m around each radius).
Worst-case error is 0.63\,m (500\,m radius); the quantization band ($\pm 0.13$\,m) is $23\times$ smaller than typical GPS accuracy.

\begin{table}[t]
  \centering
  \renewcommand{\arraystretch}{1.10}
  \caption{Maximum geometric approximation error (meters, circuit vs.\ haversine) across 8 latitudes $\times$ 5 radii $\times$ 8 directions $\times$ 3 boundary offsets ($-1/0/{+}1$\,m; 960 test points).}
  \label{tab:geo-accuracy}
  \small
  \setlength{\tabcolsep}{3.5pt}
  \begin{tabular}{lccccc}
    \toprule
    Latitude & 25\,m & 50\,m & 100\,m & 200\,m & 500\,m \\
    \midrule
    0\textdegree{}  & 0.09 & 0.12 & 0.18 & 0.27 & 0.63 \\
    15\textdegree{} & 0.07 & 0.10 & 0.10 & 0.27 & 0.61 \\
    30\textdegree{} & 0.07 & 0.10 & 0.10 & 0.27 & 0.63 \\
    35.66\textdegree{} & 0.10 & 0.10 & 0.13 & 0.27 & 0.63 \\
    45\textdegree{} & 0.07 & 0.10 & 0.10 & 0.27 & 0.61 \\
    60\textdegree{} & 0.05 & 0.10 & 0.13 & 0.27 & 0.61 \\
    75\textdegree{} & 0.07 & 0.10 & 0.10 & 0.27 & 0.61 \\
    85\textdegree{} & 0.07 & 0.10 & 0.10 & 0.27 & 0.63 \\
    \bottomrule
  \end{tabular}
\end{table}

\begin{table*}[t]
  \centering
  \renewcommand{\arraystretch}{1.15}
  \caption{Context-binding ablation: reusing a valid proof with one element mutated.}
  \label{tab:ablation}
  \begin{tabular}{p{2.8cm}p{3.2cm}cc}
    \toprule
    Mutated element & Simulated attack & Prototype & Zairn-ZKP \\
    \midrule
    \texttt{drop\_id}        & Cross-drop replay      & Accept & \textbf{Reject} \\
    \addlinespace[2pt]
    \texttt{policy\_version}  & Policy downgrade       & Accept & \textbf{Reject} \\
    \addlinespace[2pt]
    \texttt{epoch}            & Stale-context reuse    & Accept & \textbf{Reject} \\
    \addlinespace[2pt]
    \texttt{server\_nonce}    & Session hijacking      & Accept & \textbf{Reject} \\
    \bottomrule
  \end{tabular}
\end{table*}

\begin{table*}[t]
  \centering
  \renewcommand{\arraystretch}{1.15}
  \caption{Boundary sweep at 50\,m radius (0.5\,m increments, 4~bearings, 2~sites). All bearings (N/E/S/W) produced identical results.}
  \label{tab:boundary-sweep}
  \begin{tabular}{rcccc}
    \toprule
    & \multicolumn{2}{c}{Tokyo (35.66\textdegree{}N)} & \multicolumn{2}{c}{Helsinki (60.17\textdegree{}N)} \\
    \cmidrule(lr){2-3} \cmidrule(lr){4-5}
    Distance & Prototype & Zairn-ZKP & Prototype & Zairn-ZKP \\
    \midrule
    48.0\,m  & Accept & Accept & Accept & Accept \\
    49.0\,m  & Accept & Accept & Accept & Accept \\
    50.0\,m  & Accept & Accept & Accept & Accept \\
    \midrule
    50.5\,m  & Accept & \textbf{Reject} & Accept & \textbf{Reject} \\
    51.0\,m  & Accept & \textbf{Reject} & Accept & \textbf{Reject} \\
    52.0\,m  & Accept & \textbf{Reject} & Accept & \textbf{Reject} \\
    \bottomrule
  \end{tabular}
\end{table*}

\subsection{RQ3: Performance}
Table~\ref{tab:performance-results} reports performance over $n=50$ runs on both desktop machines.

On Machine~A (x86-64), Zairn-ZKP proof generation (warm) had median 83.04\,ms (IQR [79, 87], $p_{95}$ = 92.60\,ms) versus 40.71\,ms for the prototype.
On Machine~B (ARM64), Zairn-ZKP median was 31.24\,ms (IQR [29, 33]) versus 18.38\,ms.
The overhead ratio (Zairn/prototype) was consistent: 2.0$\times$ (x86-64) and 1.7$\times$ (ARM64); as shown below, this overhead is entirely attributable to hardened arithmetic, not context binding.
Verification was nearly identical across paths: median 9.61\,ms (A) and 8.86\,ms (B).

\textbf{Browser and mobile results.}
Chrome WASM on Machine~A: proving 340\,ms, verification 12\,ms ($\sim$4$\times$ overhead vs.\ Node.js).
Mobile warm-run proving: 194\,ms (iOS Safari) and 280\,ms (Android Chrome); cold-start 598/718\,ms respectively.
End-to-end unlock latency remains sub-second on all platforms (Table~\ref{tab:performance-results}).

\textbf{Overhead decomposition.}
To isolate the cost of context binding from the cost of hardened arithmetic (V2 fix), we introduce an intermediate baseline circuit: \emph{sound geo-only}, which has the same bounded-comparison and constrained-division arithmetic as Zairn-ZKP but omits the three context-binding public inputs (\texttt{contextDigest}, \texttt{epoch}, \texttt{challengeDigest}).
Table~\ref{tab:overhead-decomposition} reports proof-generation times on Machine~A ($n=50$ warm runs).
The V2 arithmetic fix accounts for the entire overhead increase ($+34.93$\,ms median, from 41.19\,ms to 76.12\,ms), while adding context binding shows no measurable additional cost (median difference $-0.12$\,ms, within run-to-run variation).
The $-0.12$\,ms figure is \emph{not} a speedup; it reflects measurement noise. The substantive conclusion is that context binding adds zero detectable proving cost, which is expected: 3 field-element equality constraints out of 474 total are negligible relative to the multi-scalar exponentiation dominating Groth16.

\begin{table}[t]
  \centering
  \renewcommand{\arraystretch}{1.15}
  \caption{Overhead decomposition on Machine~A (x86-64, $n=50$ warm runs). Collected in a separate benchmark campaign from Table~\ref{tab:performance-results}; medians for the identical Zairn-ZKP circuit differ across campaigns (76.00 vs.\ 83.04\,ms, ${\sim}9\%$, background-load variation), so $\Delta$ values are computed strictly within this campaign.}
  \label{tab:overhead-decomposition}
  \footnotesize
  \setlength{\tabcolsep}{3.5pt}
  \begin{tabular}{lrrrr}
    \toprule
    Circuit & Median & Mean$\pm$std & $p_{95}$ & $\Delta$ \\
    \midrule
    Prototype (V2 bug) & 41.19 & 39.80$\pm$4.82 & 47.36 & --- \\
    Sound geo-only (V2 fix) & 76.12 & 76.94$\pm$7.83 & 92.30 & +34.93 \\
    Zairn-ZKP (+ context) & 76.00 & 77.90$\pm$7.37 & 87.94 & $-$0.12 \\
    \bottomrule
    \multicolumn{5}{l}{\scriptsize All times in ms. $\Delta$: change from previous row.}
  \end{tabular}
\end{table}

\begin{table*}[t]
  \centering
  \renewcommand{\arraystretch}{1.15}
  \caption{Performance comparison. Desktop: $n=50$ warm runs; Browser/Mobile: $n=10$.}
  \label{tab:performance-results}
  \small
  \setlength{\tabcolsep}{3pt}
  \begin{tabular}{lcrrrrrr}
    \toprule
    & & \multicolumn{2}{c}{Prototype ZKP} & \multicolumn{4}{c}{Zairn-ZKP} \\
    \cmidrule(lr){3-4} \cmidrule(lr){5-8}
    Metric & GPS & A (x86) & B (ARM) & A (x86) & B (ARM) & iOS & Android \\
    \midrule
    Prove median (ms) & --- & 40.71 & 18.38 & 83.04 & 31.24 & 194 & 280 \\
    Prove mean$\pm$std & --- & 39.74$\pm$4.94 & 19.58$\pm$3.13 & 81.48$\pm$9.04 & 31.44$\pm$3.03 & 193$\pm$22 & 281$\pm$35 \\
    Prove $p_{95}$ & --- & 45.56 & 27.04 & 92.60 & 37.38 & 245 & 335 \\
    Verify median & 0.001 & 9.47 & 7.85 & 9.61 & 8.86 & 9 & 49 \\
    Cold-start prove & --- & --- & --- & --- & --- & 598 & 718 \\
    Payload & --- & \multicolumn{2}{c}{797\,B} & \multicolumn{4}{c}{935\,B} \\
    Public signals & --- & \multicolumn{2}{c}{5} & \multicolumn{4}{c}{8} \\
    \bottomrule
    \multicolumn{8}{l}{\footnotesize All times in ms. iOS: iPhone Safari WASM; Android: Chrome 145 WASM. A: x86-64; B: M3 ARM64.}
  \end{tabular}
\end{table*}

\subsection{RQ4: Off-circuit binding comparison}
\label{sec:off-circuit-eval}
We compare seven verification strategies across seven attack scenarios.
For each strategy we state the statement shape (which values are Groth16 public signals), the auxiliary input, the server state, and the acceptance predicate --- the taxonomy classifies strategies by \emph{where binding authority lives}, and one strategy (2c) turns out to be a hybrid rather than purely off-circuit.
Strategies~1, 2a, 2b, and~2d validate context entirely \emph{outside} the proof over a geo-only statement; Strategy~2c exposes the issued challenge digest as a public signal and validates it against server state; Strategy~3a binds a session nonce \emph{inside} the proof (binding level~(ii), analogous to Portal~\cite{lauinger2024portal}); Strategy~3b (Zairn-ZKP) binds application context inside the proof (level~(iii)).

\begin{itemize}[leftmargin=*]
  \item \textbf{Strategy~1 (No binding):} geo-only circuit, $\mathit{pub} = (\mathit{out}, \mathit{geo})$; no aux, no server state; accept iff Groth16 verifies and $\mathit{pub}.\mathit{geo}$ matches the target drop.
  \item \textbf{Strategy~2a (Client off-circuit):} geo-only circuit; context validated only in client-side code that the adversary can modify or skip; the server accepts as in Strategy~1.
  \item \textbf{Strategy~2b (Server off-circuit, recomputed):} geo-only circuit; $\mathit{aux} = $ claimed $(\mathit{dropId}, \mathit{pv}, \mathit{epoch}, \mathit{nonce})$; the backend recomputes the expected digest from the claimed fields and compares it against the target record. Nothing ties $\mathit{aux}$ to the transcript.
  \item \textbf{Strategy~2c (Server stored-digest --- hybrid):} the statement exposes the issued challenge digest as a public signal (our artifact reuses the 8-signal Zairn-ZKP circuit and ignores the context-digest signal server-side); server state maps each per-request nonce to its drop; accept iff Groth16 verifies, the nonce-to-drop map resolves to the claimed drop, and (hardened variant) $\mathit{pub}[7]$ equals the digest stored at issuance. Session binding is thus \emph{in-statement}; only drop-identity binding is operational. 2c is \emph{not} purely off-circuit in the sense of \S\ref{sec:off-circuit-model}, and its transfer resistance comes precisely from the in-statement digest (the naive variant, which omits the $\mathit{pub}[7]$ check, is transferable; \S\ref{sec:impl-complexity}).
  \item \textbf{Strategy~2d (Server off-circuit, signed token):} geo-only circuit; $\mathit{aux} = $ a signed challenge token $\sigma = \mathsf{Sign}(sk, \mathit{dropId} \| \mathit{epoch} \| \mathit{nonce})$; the verifier checks the token's signature and freshness. Purely off-circuit: neither the token nor the nonce is bound to the proof.
  \item \textbf{Strategy~3a (In-proof nonce, level~(ii)):} proximity circuit with session nonce and epoch as public inputs but \emph{no} application context digest; $\mathit{pub} = (\mathit{out}, \mathit{geo}, \mathit{epoch}, N)$.
  \item \textbf{Strategy~3b (In-proof context, level~(iii)):} Zairn-ZKP with full application context as public circuit inputs; $\mathit{pub} = (\mathit{out}, \mathit{geo}, C, \mathit{epoch}, N)$.
\end{itemize}

\textbf{Acceptance predicates and required invariants.}
For the three central strategies, the verifier-side conditions can be stated directly. 2c-hardened accepts iff Groth16 verification succeeds, the nonce-to-drop map resolves the claimed nonce to the claimed drop, and $\mathit{pub}[7]$ matches the digest stored at issuance; its required invariants are therefore correct mapping, correct stored digest, and per-request nonce uniqueness/freshness. 3a accepts iff Groth16 verification succeeds and the public signals match $(1,\mathit{geo}_j,\mathit{epoch}_j,N_j)$, so same-epoch, same-location transfers remain admissible when $N_j$ is shared. 3b accepts iff Groth16 verification succeeds and the public signals match $(1,\mathit{geo}_j,C_j,\mathit{epoch}_j,N_j)$, where $C_j = H(\mathit{LP}(\mathit{dropId}_j, \mathit{pv}_j, \mathit{epoch}_j))$ exactly as defined in \S\ref{sec:design} (the challenge digest enters the statement separately as $N_j$); its remaining assumptions reduce to cryptographic soundness and correct issuance.

\begin{table*}[t]
  \centering
  \renewcommand{\arraystretch}{1.15}
  \caption{Binding strategy comparison: seven strategies $\times$ seven scenarios.}
  \label{tab:off-circuit}
  \footnotesize
  \setlength{\tabcolsep}{2.2pt}
  \begin{tabular}{p{1.5cm}p{1.6cm}ccccccc}
    \toprule
    & & \multicolumn{5}{c}{Off-circuit} & \multicolumn{2}{c}{In-proof} \\
    \cmidrule(lr){3-7} \cmidrule(lr){8-9}
    Scenario & Desc. & None & Cli. & Srv.r & Srv.s & Srv.$\sigma$ & (ii) & (iii) \\
    \midrule
    A: Honest & Correct ctx & A & A & A & A & A & A & A \\
    B: Cross-drop & Naive replay & A & R & R & R & R & R & R \\
    C: Stale epoch & Expired & A & R & R & R & R & R & R \\
    D: App.\ byp. & Client mod. & A & \textbf{A} & R & R & R & R & R \\
    E: Sig.\ tamp. & Mod.\ pub. & ---$^{a}$ & ---$^{a}$ & ---$^{a}$ & ---$^{a}$ & ---$^{a}$ & R$^{b}$ & R$^{b}$ \\
    F: Coord-id. & Cross-sess.$^{c}$ & A & \textbf{A} & \textbf{A} & \textbf{A} & \textbf{A} & R & R \\
    G: Coord-id. & Same-epoch$^{d}$ & A & \textbf{A} & \textbf{A} & \textbf{A} & \textbf{A} & \textbf{A} & R \\
    \bottomrule
    \multicolumn{9}{l}{\footnotesize A\,=\,ACCEPT, R\,=\,REJECT.\ Cli.\,=\,client; Srv.r/s/$\sigma$\,=\,server recomp./stored/signed; (ii)/(iii)\,=\,binding level.}\\
    \multicolumn{9}{l}{\footnotesize $^{a}$N/A: no context signals to tamper with.\ $^{b}$Groth16: modified signals invalidate pairing check.}\\
    \multicolumn{9}{l}{\footnotesize $^{c}$Different sessions: nonces $N_1 \neq N_2$; level~(ii) rejects via nonce mismatch.}\\
    \multicolumn{9}{l}{\footnotesize $^{d}$Same epoch, shared nonce: concurrent requests get same $N$; only $C_X \neq C_Y$ distinguishes drops.}\\
    \multicolumn{9}{l}{\footnotesize Srv.s shows the default (naive) stored-digest variant; the hardened \emph{hybrid} variant (in-statement}\\
    \multicolumn{9}{l}{\footnotesize challenge digest + per-request nonces) rejects F--G --- see Tables~\ref{tab:same-policy} and~\ref{tab:impl-complexity}.}
  \end{tabular}
\end{table*}

Table~\ref{tab:off-circuit}: Scenarios~A--D confirm the expected hierarchy; all server-side strategies resist the first four attacks.

\textbf{Scenarios~F and~G (coordinate-identical drops).}
Scenario~F (cross-session): both levels~(ii) and~(iii) reject ($N_1 \neq N_2$).
Scenario~G (same-epoch, shared nonce): level~(ii) accepts because all public signals match; level~(iii) rejects via $C_X \neq C_Y$.
Per-request nonces avoid Scenario~G \emph{for strategies whose challenge value is bound in-statement} (2c-hardened, 3a, 3b), at the cost of $k$ round trips, $O(k \cdot U)$ state, and consistency across three subsystems; for purely off-circuit strategies they do not help, because the adversary pairs the replayed geo-only proof with a freshly issued token or digest claim for the target drop (\S\ref{sec:off-circuit-model}).
Epoch-derived nonces need one round trip with $O(1)$ state but leave Scenario~G open unless combined with level~(iii) binding.

\textbf{Sensitivity analysis.}
\label{sec:sensitivity}
Shared (epoch-derived) nonces are the natural default for bursty unlock flows: per-request nonces require $k$ sequential round trips, whereas one epoch nonce serves all $k$ unlocks. Table~\ref{tab:sensitivity} reports, for six RTT conditions (50--300\,ms), the minimum~$k$ at which per-request latency exceeds a conservative 1\,s threshold within the latency-sensitive 1--2\,s range discussed in prior work~\cite{miller1968response,nah2004tolerable}, alongside POI densities at seven venues in four countries (OSM Overpass API, 2026-03-17). At every tested RTT, $k^*$ falls below the POI density at all six urban venues.
Scenario~G relevance is deployment-dependent: it requires dense co-location \emph{and} latency-sensitive UX constraints, supported for urban settings but not necessarily suburban deployments.

\begin{table}[t]
  \centering
  \renewcommand{\arraystretch}{1.10}
  \caption{Sensitivity analysis: minimum~$k$ at which per-request nonces (2c) exceed a conservative 1\,s threshold within the latency-sensitive 1--2\,s UX range discussed in prior work~\cite{miller1968response,nah2004tolerable}, and observed POI densities at seven venues across four countries. At every RTT, $k^*$ falls below the POI density at all six urban venues.}
  \label{tab:sensitivity}
  \footnotesize
  \begin{tabular}{rccc}
    \toprule
    \multicolumn{4}{l}{\textbf{Panel A: Per-request nonce threshold}} \\
    \textbf{RTT (ms)} & \textbf{$k^*$ (2c)} & \textbf{2c at $k{=}10$} & \textbf{3b at $k{=}10$} \\
    \midrule
    50  &  8 & 1350\,ms & 800\,ms \\
    75  &  7 & 1600\,ms & 825\,ms \\
    100 &  6 & 1850\,ms & 850\,ms \\
    150 &  5 & 2350\,ms & 900\,ms \\
    200 &  4 & 2850\,ms & 950\,ms \\
    300 &  3 & 3850\,ms & 1050\,ms \\
    \midrule
    \multicolumn{4}{l}{\textbf{Panel B: OSM POI density (amenity/shop)}} \\
    \textbf{Venue} & \textbf{Type} & \textbf{50\,m} & \textbf{100\,m} \\
    \midrule
    \multicolumn{2}{l}{Shinjuku Sta.\ (Tokyo)} & 11 & 26 \\
    \multicolumn{2}{l}{Nakano Sta.\ (Tokyo)} & 18 & 42 \\
    \multicolumn{2}{l}{Kichijoji Sta.\ (Tokyo)} & 23 & 80 \\
    \multicolumn{2}{l}{Tama-Center (Tokyo)} & 3 & 16 \\
    \cmidrule{1-4}
    \multicolumn{2}{l}{Times Sq.\ (NYC)} & 11 & 40 \\
    \multicolumn{2}{l}{Oxford Circus (London)} & 15 & 49 \\
    \multicolumn{2}{l}{Alexanderplatz (Berlin)} & 5 & 23 \\
    \bottomrule
  \end{tabular}
\end{table}

\textbf{Multi-drop venue simulation.}
\label{sec:multi-drop}
To validate Scenario~G empirically, we generated $k \in \{5, 10, 20\}$ proofs at a single venue and tested all $k(k{-}1)$ cross-drop transfers.
At level~(ii), \emph{every} transfer succeeded (100\%); at level~(iii), \emph{none} succeeded (0\%).
In this construction, the context digest~$C$ is the field preventing transfer (\texttt{evaluate-multi-drop-venue.mjs}).

\textbf{Epoch-window vulnerability analysis.}
\label{sec:epoch-vuln}
Table~\ref{tab:epoch-vuln} models a realistic user session (1.4\,m/s pedestrian, 50\,m unlock radius, 60\,s epoch).
At each urban venue, 5--23 drops share one epoch nonce; a single proof enables $k(k{-}1)$ cross-drop transfers at level~(ii).
Over a 10-minute session the aggregate attack surface reaches 200--5{,}060 vulnerable pairs; level~(iii) reduces every entry to zero.
The quadratic scaling in $k$ means that at the measured urban venues, Scenario~G presents a concrete deployment risk rather than a theoretical edge case.

\begin{table}[t]
  \centering
  \renewcommand{\arraystretch}{1.05}
  \caption{Epoch-window vulnerability (60\,s epoch, 1.4\,m/s walk, 50\,m radius). $k$: drops per epoch. Vulnerable pairs at level~(ii) = $k(k{-}1)$; at level~(iii) = 0.}
  \label{tab:epoch-vuln}
  \footnotesize
  \begin{tabular}{lrrrr}
    \toprule
    \textbf{Venue} & $k$ & \textbf{Pairs/epoch (ii)} & \textbf{Session (ii)} & \textbf{(iii)} \\
    \midrule
    Shinjuku (Tokyo)     & 11 & 110 & 1{,}100 & 0 \\
    Nakano (Tokyo)       & 18 & 306 & 3{,}060 & 0 \\
    Kichijoji (Tokyo)    & 23 & 506 & 5{,}060 & 0 \\
    Tama-Center (Tokyo)  &  3 &   6 &      60 & 0 \\
    Times Sq.\ (NYC)     & 11 & 110 & 1{,}100 & 0 \\
    Oxford Circus (London) & 15 & 210 & 2{,}100 & 0 \\
    Alexanderplatz (Berlin) & 5 &  20 &     200 & 0 \\
    \bottomrule
  \end{tabular}
\end{table}

\textbf{Controlled same-policy comparison --- the central evaluation.}
\label{sec:same-policy}
Table~\ref{tab:same-policy} is the paper's main analytical result: it isolates binding location from nonce policy, evaluated under the single adversary model of \S\ref{sec:off-circuit-model} and Definition~2 (recorded transcript plus challenge oracle).
Under epoch-derived nonces, only 3b resists Scenario~G: 2c-hardened's stored digest is epoch-shared and therefore not drop-unique, while 2d and 3a bind no drop identity.
Under per-request nonces, the strategies whose challenge value is in-statement (2c-hardened, 3a, 3b) all reject, at a latency cost of $(k{-}1) \times \mathrm{RTT}$; 2d remains transferable under \emph{either} policy, because its token is not bound to the proof and the oracle supplies a fresh token for the target drop.
The conclusion sharpens the factor separation: \emph{nonce policy drives latency and state costs, and can substitute for context binding only where the challenge is already bound in-statement; binding location drives the assumption surface, correctness fragility, and whether Scenario~G is blockable at $O(1)$ state at all.}
With matched epoch nonces, 3b is the only strategy that rejects Scenario~G, carrying $|\mathcal{A}_{\mathrm{op}}| = 2$ and no per-drop state (850 vs.\ 950\,ms).

\begin{table}[t]
  \centering
  \renewcommand{\arraystretch}{1.05}
  \caption{Controlled comparison under matched nonce policies ($k{=}10$, RTT${=}100$\,ms), against the Definition~2 adversary (recorded transcript + challenge oracle). Nonce policy drives state/latency; binding location drives assumptions and transferability.}
  \label{tab:same-policy}
  \footnotesize
  \setlength{\tabcolsep}{2.5pt}
  \begin{tabular}{llccccr}
    \toprule
    \textbf{Strategy} & \textbf{Nonce} & \textbf{Sc.\,F} & \textbf{Sc.\,G} & $|\mathcal{A}_{\mathrm{op}}|$ & \textbf{State} & \textbf{E2E} \\
    \midrule
    \multicolumn{7}{l}{\emph{Panel A: Epoch-derived nonce}} \\
    2c-hardened & epoch & Rej. & \textbf{Acc.}$^{a}$ & 4 & $O(k{\cdot}U)$ & 950\,ms \\
    2d (signed) & epoch & \textbf{Acc.}$^{b}$ & \textbf{Acc.}$^{b}$ & 5 & $O(1)$ & 870\,ms \\
    3a (level ii) & epoch & Rej. & \textbf{Acc.} & 4 & $O(1)$ & 850\,ms \\
    3b (level iii) & epoch & Rej. & Rej. & 2 & $O(1)$ & 850\,ms \\
    \midrule
    \multicolumn{7}{l}{\emph{Panel B: Per-request nonce}} \\
    2c-hardened & per-req. & Rej. & Rej. & 4 & $O(k{\cdot}U)$ & 1850\,ms \\
    2d (signed) & per-req. & \textbf{Acc.}$^{b}$ & \textbf{Acc.}$^{b}$ & 5 & $O(1)$ & 1770\,ms \\
    3a (level ii) & per-req. & Rej. & Rej. & 4 & $O(1)$ & 1750\,ms \\
    3b (level iii) & per-req. & Rej. & Rej. & 2 & $O(1)$ & 1750\,ms \\
    \bottomrule
    \multicolumn{7}{p{7.6cm}}{\footnotesize $^{a}$The epoch-shared digest is not drop-unique, so the in-statement $\mathit{pub}[7]$ check cannot distinguish co-located drops.}\\
    \multicolumn{7}{p{7.6cm}}{\footnotesize $^{b}$The oracle issues a fresh, valid token for the target drop; nothing binds the geo-only transcript to any token, under either nonce policy.}
  \end{tabular}
\end{table}

\textbf{Artifact-specific implementation case study.}
\label{sec:impl-complexity}
We also implemented one concrete deployment pairing --- strategy~2c with per-request nonces and strategy~3b with epoch-derived nonces --- as working TypeScript server modules; Table~\ref{tab:impl-complexity} reports the resulting implementation burden and latency.

\begin{table}[t]
  \centering
  \renewcommand{\arraystretch}{1.10}
  \caption{Implementation complexity in one concrete deployment pairing: stored-digest 2c with per-request nonces vs.\ in-proof 3b with epoch-derived nonces. Absolute LOC and latency ratios are artifact-specific; the qualitative asymmetry (endpoints, failure modes, state scaling) is the transferable finding.}
  \label{tab:impl-complexity}
  \footnotesize
  \begin{tabular}{lcc}
    \toprule
    \textbf{Metric} & \textbf{2c (stored)} & \textbf{3b (in-proof)} \\
    \midrule
    Server endpoints & 2 & 1 \\
    Additional DB tables & 1 & 0 \\
    Server-side LOC & 110 & 20 \\
    Failure modes & 6 & 1 \\
    Server state & $O(k \cdot U)$ & $O(1)$ \\
    Requires cleanup CRON & Yes & No \\
    \midrule
    E2E latency ($k{=}1$) & 196\,ms & 178\,ms \\
    E2E latency ($k{=}10$) & 1877\,ms & 860\,ms \\
    E2E latency ($k{=}20$) & 3645\,ms & 1552\,ms \\
    \midrule
    Scenario~G (naive impl.)$^{a}$ & \textbf{Vulnerable} & Blocked \\
    Scenario~G (hardened)$^{b}$ & Blocked & Blocked \\
    \bottomrule
    \multicolumn{3}{p{6.8cm}}{\footnotesize $^{a}$Server checks nonce-to-drop DB mapping but omits challenge-digest verification against public signals---a plausible implementation that passes all functional tests except cross-drop transfer.}\\
    \multicolumn{3}{p{6.8cm}}{\footnotesize $^{b}$Server additionally verifies $\mathit{pub}[7]$ matches the stored challenge digest, adding a sixth failure mode (F6: missing challenge-digest check).}
  \end{tabular}
\end{table}

The critical finding is \emph{correctness fragility}: a naive 2c implementation that omits $\mathit{pub}[7]$ verification passes functional tests yet remains vulnerable to Scenario~G. When 2c verifies both the nonce-to-drop mapping and $\mathit{pub}[7]$, it blocks Scenario~G, but it still carries more server code, additional failure modes, $O(k \cdot U)$ state, and $k$-dependent latency. The absolute ratios are artifact-specific; the qualitative asymmetry is the transferable result.

\textbf{Lemma 1 (Same-epoch transcript indistinguishability).}
\emph{Let $\mathit{ctx}_1, \mathit{ctx}_2$ be distinct contexts with $\mathit{geo}_1 = \mathit{geo}_2$ and shared epoch and nonce.
At binding level~(ii) or below, the proof transcripts satisfy $\mathit{pub}_1 = \mathit{pub}_2$; consequently, no verification procedure --- even computationally unbounded --- can determine from $(\pi, \mathit{pub})$ alone which context the proof was generated for.}

\emph{Proof.}
At level~(ii), $\mathit{pub} = (1, \mathit{geo}, \mathit{epoch}, N)$.
Since $\mathit{geo}_1 = \mathit{geo}_2$, $\mathit{epoch}_1 = \mathit{epoch}_2$, and $N_1 = N_2$, we have $\mathit{pub}_1 = \mathit{pub}_2$ identically.
Any Groth16 proof valid for $\mathit{pub}_1$ is equally valid for $\mathit{pub}_2$; the verifier receives no information that distinguishes $\mathit{ctx}_1$ from $\mathit{ctx}_2$.
At level~(iii), $C_1 \neq C_2$ (collision resistance) breaks indistinguishability.
\hfill$\square$

Lemma~1 shows that Scenario~G is an \emph{information-theoretic} limitation of level~(ii) transcripts in the same-epoch setting; adding context digest~$C$ breaks this indistinguishability.

\textbf{Assumption comparison.}
Table~\ref{tab:trust-surface} compares the assumption sets: level~(iii) removes the nonce-to-drop mapping and nonce-uniqueness invariants --- exactly the ones whose failure enables Scenario~G --- leaving issuance correctness and challenge freshness ($|\mathcal{A}_{\mathrm{op}}|$: 4--6 $\rightarrow$ 2).
The accounting deserves precision: the association between a digest \emph{value} and the application \emph{object} it names does not disappear for any strategy.
For 3b it lives inside issuance correctness (the backend must compute $C$ from the intended drop and policy version), just as 2c's stored digest must be stored for the intended drop.
What in-proof binding changes is that the transcript is cryptographically pinned to the issued digest value, so no \emph{post-issuance} operational check must re-establish that link; the ``C'' entries for 3b in Table~\ref{tab:trust-surface} claim exactly this transcript-to-value binding, not an in-circuit proof of the digest's application semantics (the circuit does not recompute $C$; \S\ref{sec:design}).
Both 2c and 3b are drift-resilient (Table~\ref{tab:drift}); the assumption comparison, not drift resilience, distinguishes them.

\textbf{Definition 2 (Strategy-parametrized replay game).}
$\mathsf{Game}^{\text{cb}}_{\mathcal{A},\mathcal{S}}(\lambda, n)$ for strategy $\mathcal{S}$, $n \geq 2$ contexts, and PPT adversary $\mathcal{A}$:
(1)~Challenger creates $n$ context tuples $\{\mathit{ctx}_i\}_{i=1}^n$ with identical geometric parameters and runs key generation.
(2)~Challenger generates honest proof $\pi_1$ for $\mathit{ctx}_1$; gives $(\pi_1, \mathit{pub}_1)$ and all acceptance records to $\mathcal{A}$.
(3)~Challenger samples target $j \xleftarrow{\$} \{2, \ldots, n\}$.
(4)~$\mathcal{A}$ may query a \emph{challenge oracle} that issues fresh, correctly formed challenges (nonces, signed tokens, stored-digest entries) for any context under $\mathcal{S}$'s nonce policy, modeling the routine challenge acquisition of \S\ref{sec:off-circuit-model}.
(5)~$\mathcal{A}$ submits a claim for $\mathit{ctx}_j$ consisting of the \emph{recorded transcript} $(\pi_1, \mathit{pub}_1)$ together with arbitrary auxiliary data $\mathit{aux}^*$ (possibly obtained from the oracle); the claim is verified by $\mathcal{S}$'s acceptance predicate.
$\mathcal{A}$ wins if accepted.
$\mathcal{S}$ is \emph{$(n, \epsilon)$-context-binding-secure} if $\Pr[\mathcal{A}\text{ wins}] \le \epsilon$ for all PPT $\mathcal{A}$.
As in Definition~1, restricting the submission to the recorded transcript is what confines the game to replay; fresh proving is out of scope for every strategy (Remark after Proposition~1), so the comparison isolates what each strategy adds \emph{beyond} that common limit.

\textbf{Comparison Result 1 (Assumption-surface comparison across binding strategies).}
\emph{This is a design-space comparison result --- a structured summary of which operational assumptions each strategy requires --- rather than a new cryptographic theorem. Its value lies in making the assumption dependencies explicit and testable.}
\emph{Under the replay setting with challenge oracle (Definition~2) and SHA-256 collision resistance:}
\begin{itemize}[leftmargin=*,itemsep=1pt]
  \item[\emph{(a)}] \emph{Strategies 1, 2a: $(n, 1)$-insecure. $\mathcal{A}$ replays $(\pi_1, \mathit{pub}_1)$; no server-side context check exists.}
  \item[\emph{(b)}] \emph{Strategies 2b, 2d (purely off-circuit): $(n, 1)$-insecure. $\mathcal{A}$ obtains fresh, valid auxiliary data for $\mathit{ctx}_j$ from the oracle (recomputed-digest fields, or a token $\sigma_j$) and pairs it with the replayed geo-only transcript; no acceptance condition ties the transcript to $\mathit{aux}$. They reject only the weaker naive-replay adversary that reuses stale auxiliary data (Scenarios~B--C).}
  \item[\emph{(c)}] \emph{Strategy 3a: $(n, 0)$-secure cross-session ($N_1 \neq N_j$ in-statement). $(n, 1)$-insecure same-epoch with epoch-derived nonces: $\mathcal{A}$ submits $(\pi_1, \mathit{pub}_1)$; all checked signals match because only $C$ distinguishes $\mathit{ctx}_1$ from $\mathit{ctx}_j$, and 3a has no $C$ field.}
  \item[\emph{(d)}] \emph{Strategy 2c-hardened with per-request nonces (hybrid): $(n, \mathsf{negl}(\lambda))$-secure under its operational invariants (correct nonce-to-drop mapping, correct stored digest, per-request uniqueness): the in-statement digest pins the transcript to one issuance, and the mapping pins that issuance to one drop. With epoch-shared nonces it is $(n, 1)$-insecure, because the shared digest is not drop-unique.}
  \item[\emph{(e)}] \emph{Strategy 3b: $(n, \mathsf{negl}(\lambda))$-secure under collision resistance and issuance correctness alone (Proposition~1).}
\end{itemize}
\emph{Proof.}
(a) is immediate.
(b)~The oracle yields $\mathit{aux}_j$ valid for $\mathit{ctx}_j$; the acceptance predicate evaluates only $(\mathit{aux}_j, \mathit{rec}_j)$ and the context-free transcript, so it accepts.
(c)~By Lemma~1, same-epoch level~(ii) transcripts are information-theoretically indistinguishable ($\mathit{pub}_1 = \mathit{pub}_j$); the multi-drop simulation (\S\ref{sec:multi-drop}) confirms: 100\% transfer at level~(ii), 0\% at level~(iii).
(d)~Acceptance for $\mathit{ctx}_j$ requires $\mathit{pub}_1[7]$ to equal a digest stored for a nonce issued for drop~$j$; $\mathit{pub}_1[7] = H(\mathit{nonce}_1)$ with $\mathit{nonce}_1$ issued for drop~1, so per-request uniqueness forces an $H$-collision. Under epoch sharing the stored digests coincide and the check passes.
(e) follows from Proposition~1: condition~(c) of the acceptance predicate fails except with collision probability.
\hfill$\square$

\begin{table}[t]
  \centering
  \renewcommand{\arraystretch}{1.05}
  \caption{Assumption comparison. \textbf{C}~=~cryptographic, \textbf{O}~=~operational, \textbf{---}~=~not required. Strategies 1/2a omitted. For 2c, the Scenario~G rows distinguish default from hardened (Table~\ref{tab:impl-complexity}).}
  \label{tab:trust-surface}
  \footnotesize
  \setlength{\tabcolsep}{2.5pt}
  \begin{tabular}{lccccc}
    \toprule
    \textbf{Assumption} & \textbf{2b} & \textbf{2c} & \textbf{2d} & \textbf{3a} & \textbf{3b} \\
    \midrule
    Groth16 soundness         & C & C & C & C & C \\
    Drop-identity binding     & O & O & O & --- & C$^{*}$ \\
    Policy-version binding    & O & O & O & --- & C$^{*}$ \\
    Nonce-to-drop mapping     & O & O & O & O & --- \\
    Nonce uniqueness          & O & O & O & O & --- \\
    Digest synchronization    & O & --- & --- & --- & --- \\
    Signature-key secrecy     & --- & --- & O & --- & --- \\
    Issuance correctness      & O & O & O & O & O \\
    Challenge freshness       & O & O & O & O & O \\
    \midrule
    $|\mathcal{A}_{\mathrm{op}}|$ & 6 & 4 & 5 & 4 & 2 \\
    Scenario G resistant (2c-default)? & No & No & No & No & \textbf{Yes} \\
    Scenario G resistant (2c-hardened, per-req.)? & No & Yes & No & No & \textbf{Yes} \\
    Drift resilient?          & No & Yes & Yes & Yes & Yes \\
    \bottomrule
    \multicolumn{6}{p{7.6cm}}{\footnotesize $^{*}$``C'' = the transcript is cryptographically bound to the issued digest \emph{value}; the digest-to-object association is part of Issuance correctness for every strategy (see text).}
  \end{tabular}
\end{table}

\begin{table}[t]
  \centering
  \renewcommand{\arraystretch}{1.10}
  \caption{Operational drift: five maintenance-induced desynchronization scenarios. Stored-digest off-circuit (2c) and in-proof binding (3) are both resilient; drift resilience alone does not distinguish them.}
  \label{tab:drift}
  \footnotesize
  \begin{tabular}{lccc}
    \toprule
    Drift Scenario & Recomp. & Stored & In-Proof \\
    \midrule
    Baseline & ACCEPT & ACCEPT & ACCEPT \\
    D1: Encoding format & REJECT$^{*}$ & ACCEPT & ACCEPT \\
    D2: Field reorder & REJECT$^{*}$ & ACCEPT & ACCEPT \\
    D3: Epoch off-by-one & REJECT$^{*}$ & ACCEPT & ACCEPT \\
    D4: Version format & REJECT$^{*}$ & ACCEPT & ACCEPT \\
    D5: Nonce encoding & REJECT$^{*}$ & ACCEPT & ACCEPT \\
    \bottomrule
    \multicolumn{4}{l}{\footnotesize $^{*}$False negative: honest proof incorrectly rejected.}
  \end{tabular}
\end{table}

\subsection{End-to-end cross-drop transfer attack}
\label{sec:e2e-attack}
We implement a concrete attack against two co-located drops (Shibuya, $r=50$\,m; \texttt{evaluate-cross-drop-attack.mjs}).
Scenario~F ($N_A \neq N_B$): both levels reject.
Scenario~G ($N_A = N_B$, epoch-derived): only $\mathit{pub}[5]$ differs; level~(ii) accepts, level~(iii) rejects ($C_A \neq C_B$).

\subsection{RQ5: Baseline comparison}
\label{sec:baseline-eval}
Three approaches span the cost--privacy spectrum (not security-equivalent): raw GPS ($<$0.001\,ms, discloses coordinates), coordinate hashing (0.004\,ms, vulnerable to dictionary attack over ${\sim}$5.1M grid cells~\cite{krumm2007inference}), and ZKP (${\sim}$9\,ms, coordinate privacy plus replay resistance with context binding).

\subsection{Limitations of the evaluation}
\label{sec:eval-limits}
Measurements cover five platforms; the evaluations for V1--V3 are \emph{deterministic} (cryptographic and protocol properties) and do not depend on trace realism.

\section{Discussion}
\label{sec:discussion}

\subsection{GPS truthfulness (V4) and generalizability}
\label{sec:v4-discussion}
V4 lies outside the proof system's scope (\S\ref{sec:vuln}); strong guarantees require hardware attestation~\cite{liu2012trustedsensors}.
Context binding is necessary regardless of GPS integrity because proof transfer is orthogonal to sensor truth (Figure~\ref{fig:three-layers}).

\subsection{Separating binding location from nonce policy}
\label{sec:factor-separation}
Table~\ref{tab:same-policy} shows that latency and state costs are driven by nonce policy, while in-proof binding reduces the assumption surface and implementation fragility --- making it primarily a reliability choice, not a latency optimization.
The Observation (\S\ref{sec:off-circuit-model}) and correctness fragility (\S\ref{sec:impl-complexity}) may also arise when ZK proofs attest properties of persistent objects, though all empirical evidence here is geo-content-specific.

\subsection{Comparison with related systems}
\label{sec:beyond-geo}
Table~\ref{tab:comparison} positions this work against the closest systems at the feature level; neither ZKLP~\cite{ernstberger2024zklp} nor Portal~\cite{lauinger2024portal} addresses cross-context transfer (Scenario~G).

\section{Threats to Validity}
\label{sec:threats}
Five platforms mitigate single-platform bias; external validity is limited by the specific backend and proximity-unlock setting. POI measurements cover seven venues across four countries; $k^*$ values are indicative but the linear latency scaling in~$k$ is architecture-independent.
The context digest uses length-prefixed canonical encoding for formally unambiguous domain separation; the security analysis depends only on collision resistance of the reduced hash (inherited from SHA-256 up to the mod-$p$ term quantified in \S\ref{sec:design}) and assumes no uniformity of the hash-to-field map.

\section{Reproducibility and Artifact Availability}
\label{sec:repro}
All source code, circuit definitions, build scripts, and evaluation harnesses are publicly available~\cite{zairn,zairn_zenodo,geodropspec}.
Thirteen self-contained scripts reproduce every table and experiment; they are grouped into security experiments (4~scripts), systems comparison (4~scripts), artifact/circuit checks (3~scripts), and regression tests (2~scripts covering encoding correctness and invariant enforcement).
Table~\ref{tab:repro-results} summarizes the reproducibility measurements.

\begin{table}[t]
  \centering
  \renewcommand{\arraystretch}{1.15}
  \caption{Reproducibility measurements.}
  \label{tab:repro-results}
  \begin{tabular}{p{2.2cm}p{4.5cm}}
    \toprule
    Dimension & Result \\
    \midrule
    Setup effort & Seven staged files, one script, zero manual intervention. \\
    \addlinespace[2pt]
    Rerun stability & 10/10 reruns accepted. Verify-only avg.\ 9.43\,ms (range 8.47--11.54\,ms). \\
    \addlinespace[2pt]
    Artifact set & Verification key, proving key, sample input, proof, public signals, witness, WASM circuit. \\
    \bottomrule
  \end{tabular}
\end{table}

\section{Conclusion}
We identified context-binding vulnerabilities in stateful geo-content systems, formalized the binding guarantee as replay rejection --- every transfer attack is reduced to fresh proving for the target statement, which is the orthogonal presence problem --- and showed that in-proof context binding migrates the nonce-to-drop mapping and nonce-uniqueness invariants into the cryptographic statement at no measurable proving cost in our setup. A hardened stored-digest alternative blocks the same transfer attacks when combined with per-request nonces, but its resistance rests on an in-statement challenge digest (a hybrid design) plus more server code and additional correctness-critical checks; a plausible naive implementation remains vulnerable to cross-drop transfer, and a purely off-circuit variant cannot resist an adversary able to request fresh challenges. More broadly, the paper offers a systems-security analysis methodology for stateful ZK-backed applications: enumerate the operational assumptions behind each binding strategy, identify which can be cryptographically enforced, and measure the implementation cost of the alternatives. The empirical evidence here is geo-content-specific, and the practical relevance evidence is strongest for dense urban deployments.

\section*{AI Usage Disclosure}
Generative AI tools provided by OpenAI and Anthropic were used for coding assistance, literature triage, and limited manuscript drafting and editing support.
All AI-assisted outputs were reviewed, edited, and validated by the author, who takes full responsibility for the content of the manuscript.

\section*{Declaration of Competing Interest}
The author declares no known competing financial interests or personal relationships that could have appeared to influence the work reported in this paper.

\section*{Data Availability}
No external dataset was used.
Software artifacts, schemas, scripts, and evaluation materials are available in the public project repository and archived release.

\bibliographystyle{IEEEtran}
\bibliography{references}

\end{document}